\documentclass[10pt, onecolumn, conference]{IEEEtran}

\usepackage{graphicx,url,tabularx,booktabs}
\usepackage[utf8]{inputenc}  
\usepackage{xspace}
\usepackage{makecell}
\usepackage{multirow}
\usepackage[table]{xcolor}
\usepackage{array}

\title{IoTEdu: Access Control, Detection, and Automatic Incident Response in Academic IoT Networks}

\author{
\IEEEauthorblockN{
    Joner Assolin\IEEEauthorrefmark{1}\IEEEauthorrefmark{2},
    Diego Kreutz\IEEEauthorrefmark{2},
    Leandro Bertholdo\IEEEauthorrefmark{3}
}
\IEEEauthorblockA{\IEEEauthorrefmark{1}IComp, Universidade Federal do Amazonas (UFAM)}
\IEEEauthorblockA{\IEEEauthorrefmark{2}AI Horizon Labs, Programa de Pós-Graduação em Engenharia de Software (PPGES)\\ Universidade Federal do Pampa (UNIPAMPA)}
\IEEEauthorblockA{\IEEEauthorrefmark{3}Universidade Federal do Rio Grande do Sul (UFRGS)}
}

\begin{document} 

\maketitle
\begin{abstract}
The growing presence of IoT devices in academic environments has increased operational complexity and exposed security weaknesses, especially in academic institutions without unified policies for registration, monitoring, and incident response involving IoT. This work presents \textit{IoTEdu}, an integrated platform that combines access control, incident detection, and automatic blocking of IoT devices. The solution was evaluated in a controlled environment with simulated attacks, achieving an average time of 28.6 seconds between detection and blocking. The results show a reduction in manual intervention, standardization of responses, and unification of the processes of registration, monitoring, and incident response.
\end{abstract}


\section{Introduction}
\label{sec:introducao}

The expansion of Internet of Things (IoT) devices in academic institutions has transformed teaching, research, and infrastructure environments into complex cyber-physical ecosystems. Sensors, actuators, and connected equipment are used in laboratory activities, continuous data collection, and building automation, expanding the attack surface and demanding coordinated monitoring mechanisms \cite{yang2024characterizing}.

Despite the growth of IoT in the country, the Brazilian scenario still presents significant weaknesses. At universities such as UFRGS and Unicamp, internal processes for enabling and managing devices involve manual and bureaucratic steps, resulting in long processing times, reaching up to six business days in the case of UFRGS\footnote{\url{https://www1.ufrgs.br/CatalogoServicos/servicos/servico?servico=3239}}
 and up to two business days for the first response at Unicamp, with no defined deadline for completion\footnote{\url{https://detic.unicamp.br/servicos/rede-sem-fio-unicamp-iot-wi-fi/}}
. In contrast, institutions such as UNIPAMPA and UFAM do not provide formal workflows, policies, or SLAs for IoT devices. This lack of standardization leads to improvised practices, operational inconsistencies, and increased exposure to vulnerabilities, especially on campuses with hundreds of devices without centralized management, reinforcing the need for control and monitoring models adapted to the national context.

Currently, there is a lack of an integrated solution that unifies access control, continuous monitoring, and automated response for Internet of Things networks in the Brazilian academic context. In order to fill this gap, the development of the IoTEdu platform~\footnote{\url{https://github.com/GT-IoTEdu/API_ERRC25}} is proposed, which consists of an integrated system for managing and monitoring the security of IoT devices. The platform incorporates adaptive policies, automation, access control, and incident response mechanisms. 

The solution was designed to interoperate with existing authentication infrastructures, reduce delays in the onboarding process, standardize institutional workflows, and strengthen operational security in academic environments. Its contributions include a unified architecture suited to the Brazilian context, a hybrid mechanism for proactive and reactive control, a simplified device onboarding process, and an automated policy framework capable of detecting, responding to, and mitigating anomalous behaviors. Unlike existing solutions, IoTEdu unifies local network control, institutional authentication, and automatic response, composing an end-to-end security workflow.

\section{Related Work}
\label{sec:trabalhosrelacionados}

Solutions aimed at the security and management of IoT devices present distinct approaches, ranging from broad commercial platforms to academic initiatives with a specific scope. Table \ref{tab:comparacao_solucoes} summarizes these approaches into categories that reflect their different focuses and capabilities. The analyzed solutions are primarily centered on device management, traffic monitoring, or access control, each addressing only part of the security lifecycle. However, none of them offers integrated mechanisms for automatic mitigation based on events captured locally in the network, which hinders rapid response in academic environments, which are typically distributed and heterogeneous.

\begin{table}[!htp]
\centering
\renewcommand{\arraystretch}{1.2}
\caption{IoT Device Management Solutions in Academic Networks.}
\label{tab:comparacao_solucoes}

\resizebox{\textwidth}{!}{
\begin{tabularx}{1.55\textwidth}{@{} p{2cm} p{3cm} p{2cm} p{9cm} X @{}}
\toprule
\textbf{Category} & \textbf{Solution} & \textbf{License} & \textbf{Main Focus} & \textbf{Limitation} \\
\midrule

IoT Management Platforms
& AWS IoT Core & Proprietary & Connectivity, device management, and IoT data analytics. & Security focused on cloud–device communication; does not address local network. \\
& Azure IoT Hub & Proprietary & Device lifecycle management and message exchange. & Local network security treated as a secondary concern. \\
& ThingsBoard & Open Source & Modular platform for monitoring and integrating IoT devices. & Security and access control have limited scope. \\
& ThingWorx / Bosch IoT Suite & Proprietary & Development of industrial applications and IoT connectivity. & Network security is not a primary focus. \\
\midrule

Academic and Educational Approaches
& SURF  & Open Source & IoT ecosystem focused on collaborative research and education. & Lack of proactive monitoring. \\
& \textit{In-house} Solutions & Open Source & Manually developed networks and device registries built internally by institutions. & Low automation and reactive security. \\
& IoTEdu & Open Source & Security management for IoT devices in academic environments with automatic threat detection, automatic blocking, and integration with pfSense and Zeek. & Dependence on specific infrastructure (pfSense, Zeek) and initial validation in a reduced-scale test environment. \\
\bottomrule
\end{tabularx}
}
\end{table}

Established platforms such as AWS IoT Core, Azure IoT Hub, and ThingsBoard offer advanced management features, secure device‑to‑cloud communication, and large‑scale processing \cite{gill2021iotplatforms}, but they treat local network security as an external responsibility, failing to address east‑west risks, traffic segmentation, or granular access control—limitations that are particularly relevant in heterogeneous academic environments. Initiatives such as the SURF platform \cite{poortinga2021surf} advance interoperability and data federation, but lack continuous monitoring, adaptive policies, and automatic response. The absence of continuous monitoring undermines the ability to respond in environments with highly dynamic traffic, which is typical of laboratories and academic networks. Institutional in‑house solutions \cite{santos2019challenges} still rely on manual registrations and open networks, resulting in fragile architectures without proactive threat detection \cite{ghimire2021security}. This set of limitations highlights the need for unified platforms that integrate access control, automation, monitoring, and mitigation, especially in the Brazilian context.  
Unlike the solutions above, IoTEdu consolidates, in a single integrated platform, IoT device management, automatic threat detection through network traffic analysis, and proactive blocking via native firewall integration, specifically designed to meet the security, governance, and scalability demands of national academic networks.

\section{Arquitetura do IoTEdu}
\label{sec:arquitetura}

\begin{figure}[!htbp]
\centering
\includegraphics[width=\textwidth]{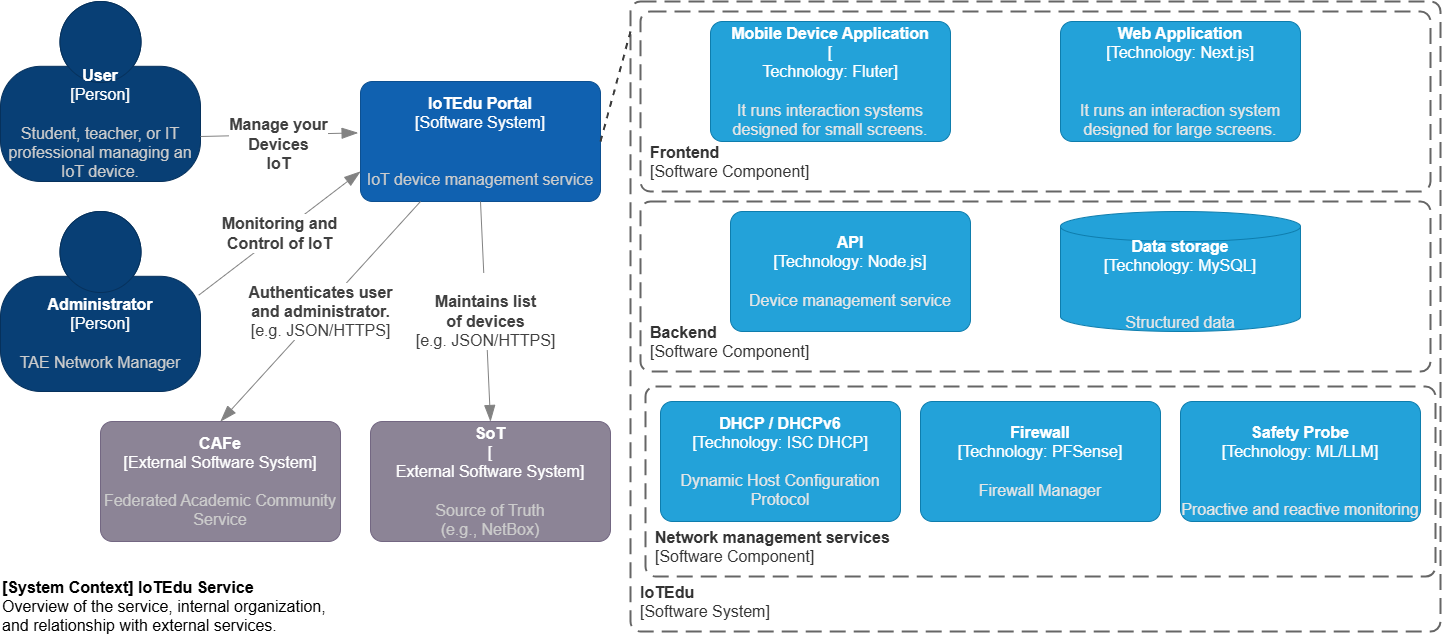}
\caption{General architecture of IoTEdu.}
\label{fig:componentes}
\end{figure}

Figure~\ref{fig:componentes} presents the overall organization of the IoTEdu architecture, structured into three layers: frontend, backend, and network management services. These components work in an integrated manner to provide registration, access control, continuous monitoring, and automatic incident response.

The platform frontend was developed with \textit{Next.js, React, and TypeScript}, adopting a component-based structure that favors scalability and continuous maintenance, enabling rapid interface evolution, an essential characteristic in multi-institutional environments. The interface dynamically adapts to each user’s profile: regular users manage their own devices; administrators control devices across the entire campus, analyze incidents, and apply blocks; and superusers manage multiple institutions and their network configurations. Communication with the backend occurs through authenticated calls to REST APIs.

The backend was built with FastAPI 2.0, following a modular architecture organized into routers for authentication, devices, \textit{firewall}, and incidents. The platform uses \textit{MySQL} with \textit{SQLAlchemy ORM} and \textit{Pydantic} models for validation, in addition to hybrid authentication based on \textit{OAuth2, SAML (CAFe), and JWT}.  
The business logic services encapsulate operations for DHCP mapping, synchronization with \textit{pfSense}, and incident correlation with network data, ensuring functional isolation, consistency across systems, and reproducibility of security actions.

Integration with \textit{pfSense} abstracts away the complexity of \textit{firewall} configuration. The platform translates block and allow actions into automated operations on aliases, filtering rules, and DHCP mappings through the \textit{pfSense} REST API. The system maintains bidirectional synchronization between the local state and the \textit{firewall} state, applying changes immediately and detecting conflicts when necessary.

The analysis module with \textit{Zeek} performs continuous log ingestion to identify suspicious events, correlating IP addresses with registered devices and classifying incidents by severity. Critical events automatically trigger the blocking mechanism in \textit{pfSense}, closing the detection and response loop within seconds. The module maintains a complete history of incidents, enabling auditing, forensic analysis, and institutional security metrics.

This integrated architecture enables IoTEdu to unify management, monitoring, and response, providing an end‑to‑end security flow tailored to the multi-institutional context of Brazilian academic networks.

\section{Evaluation and Testing}
\label{sec:metodologia}
This section describes the experimental environment, the methodology used for system validation, and the quantitative analysis of performance and effectiveness results.

\subsection{Test Environment}

The test environment was fully virtualized on a single notebook, using VirtualBox for the network infrastructure and Docker for auxiliary services. The virtual machine ran pfSense as the gateway, firewall, DHCP/DNS server, and Wi-Fi access point of an isolated network, while Zeek monitored all traffic on the virtual LAN. The IoTEdu API was executed on the physical host, receiving alerts from Zeek and sending blocking commands to pfSense. The host was equipped with an Intel Core i7-1185G7, 32 GB of RAM, and a Wi-Fi 6 interface, and the VM was configured with pfSense 2.8.1-RELEASE, 4 vCPUs, 8 GB of RAM, and Zeek 3.0.6\_6 for threat detection.

\subsection{Testing Methodology}
\label{subsec:metodologia_testes}

\begin{figure}[!htbp]
\centering
\includegraphics[width=\textwidth]{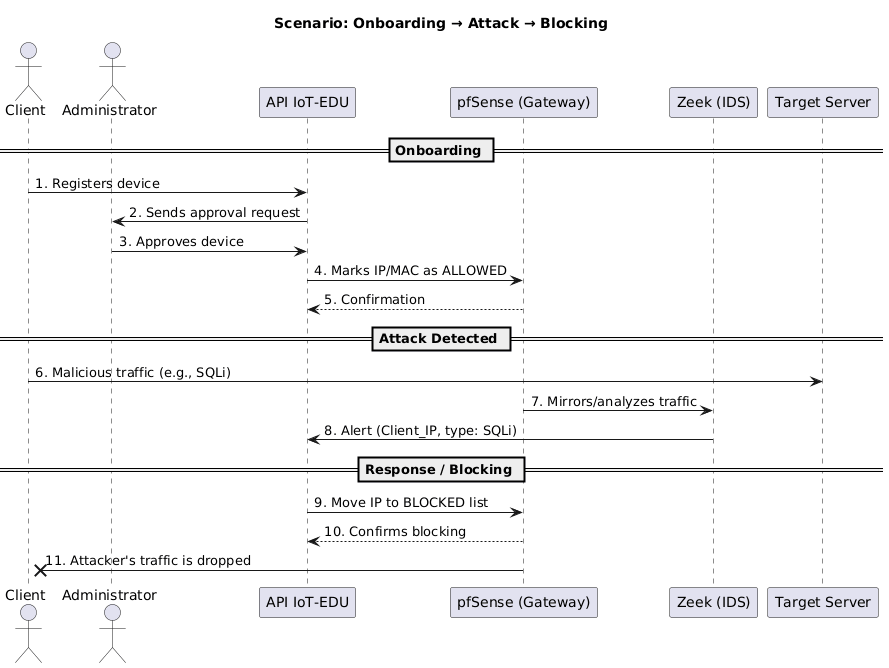}
\caption{Test Sequence Diagram}
\label{fig:sequencia_teste}
\end{figure}

The system validation followed a complete operational flow, divided into two main stages as illustrated in Figure~\ref{fig:sequencia_teste}. In the initial phase, focused on provisioning and authorization, the client authenticates in the IoTEdu API, requests access, and waits for the administrator’s approval, who validates the device and authorizes its entry into the network. Next, the API instructs pfSense to register the equipment’s MAC and IP addresses in the list of allowed devices, ensuring its controlled integration into the test environment.

Subsequently, in the second phase, corresponding to the incident detection and response cycle, an attack is simulated to assess the system’s capability for automatic identification and mitigation. The client launches an SQL Injection attack against a target server using the \textit{sqlmap}\footnote{\url{https://sqlmap.org}} tool, and the exact start time of the attack is recorded for measurement purposes. Zeek, in its continuous monitoring, identifies the malicious payload and generates an alert in the \texttt{notice.log} file. This alert is collected and processed by the \textit{ZeekService}, which extracts the relevant data and records them in the database through the \textit{IncidentService} in the \texttt{zeek\_incidents} table. After incident validation, the IoTEdu API executes the automatic response: it triggers pfSense, via the \textit{pfSenseClient}, to remove the device from the allowed list and add it to the blocked list, recording this moment in the \texttt{blocking\_feedback\_history} table. From that operation onward, the client immediately loses access to the network. To confirm the action, an independent \textit{curl}-based script monitors connectivity and records the effective disconnection time, allowing accurate measurement of the total latency of the detection and response cycle.

\subsection{Incident Response Evaluation}
\label{subsec:avaliacao_resposta}

Ten tests were conducted to evaluate the system’s Time to Detect (TtD) and Time to Block (TtB). The results, presented in Table \ref{tab:metricas_tempo}, showed an average TtD of 12.1 seconds by Zeek, with low variation. The API processing took 16.8 seconds, representing the main bottleneck. The block on the firewall was applied in 1.0 second, with loss of access occurring after an additional 1.4 seconds. The system’s total response time, from the start of the attack until its neutralization, was 28.6 seconds, validating the effectiveness of the automated detection and mitigation cycle.

\begin{table}[h!]
\centering
\renewcommand{\arraystretch}{1.3}
\caption{Average Response Time}
\label{tab:metricas_tempo}

\resizebox{\textwidth}{!}{
\begin{tabularx}{1.1\textwidth}{@{} p{5.5cm} p{1.2cm} X @{}}
\toprule
\textbf{Metric} & \textbf{Value} & \textbf{Notes} \\
\midrule
Detection Time (TtD) & 12.1 s & Zeek on the pfSense VM; variation between 11 and 14 s \\
API Processing Time & 16.8 s & Main bottleneck; involves six sequential operations \\
Blocking Time (TtB) & 1.0 s & Immediate application of rules via the pfSense API \\
Time Until Loss of Access & 1.4 s & Direct effect of the firewall; low variability \\
Total Response Time & 28.6 s & Influenced by virtualization and serial operations \\
\bottomrule
\end{tabularx}
}
\end{table}

\subsection{Performance Analysis}

The performance analysis shows that the virtualized environment directly impacts the system’s response time, with latencies between the \textit{host} and \textit{pfSense} ranging from 1 ms to 237 ms (an average of 25 ms), and the Wi-Fi link fluctuating between 2 ms and 327 ms. This combination of network variability and virtualization \textit{overhead} creates temporal instability and increases the accumulated delay during the sequential execution of the API operations. As summarized in Table~\ref{tab:analise_geral}, the incident processing stages (3.222 s) and the interactions with \textit{pfSense} (4.559 s in total) were the most costly, reflecting sensitivity to network latency and I/O operations. The system used 214.86 MB of RAM, with stable usage, and the observed CPU load corresponds to the natural accumulation of internal operations, indicating that performance is mainly limited by communication rather than computational capacity. With an average time of 2.468 s per operation, the results show that optimizations should prioritize latency reduction and improvement of the interactions among \textit{API, pfSense, and Zeek}, since the local \textit{hardware} does not represent a significant bottleneck.

\begin{table}[h!]
\centering
\renewcommand{\arraystretch}{1.3}
\caption{Execution Time and Resource Consumption per API Operation}
\label{tab:analise_geral}
\resizebox{0.9\textwidth}{!}{
\begin{tabular}{p{6cm}ccc}
\toprule
\textbf{Operation} & \textbf{Time (s)} & \textbf{CPU (\%)} & \textbf{RAM (MB)} \\
\midrule
ZeekService.get\_logs & 2.281 & 6.80 & 35.79 \\
IncidentService.save\_incident & 2.499 & 13.70 & 35.79 \\
IncidentService.process\_incidents & 3.222 & 16.10 & 35.79 \\
AliasService.get\_alias\_by\_name & 2.245 & 16.70 & 35.79 \\
AliasService.add\_addresses\_to\_alias & 2.233 & 25.00 & 35.85 \\
pfsense\_client.apply\_changes\_firewall & 2.326 & 38.90 & 35.85 \\
\hline
\textbf{Overall Total} & \textbf{14.806} & \textbf{117.20} & \textbf{214.86} \\
\bottomrule
\end{tabular}
}
\end{table}

\section{Final Considerations}
\label{sec:conclusao}

The system proved effective in controlling IoT devices and in automatically responding to incidents by integrating \textit{pfSense, Zeek} and the IoTEdu API into a complete detection and blocking workflow, achieving an average response time of 28.6 seconds. As a natural evolution, the platform can incorporate machine-learning-based detection, device \textit{fingerprinting}, and integration with IDS/IPS, increasing the accuracy of the automatic response. The modular architecture and profile-based \textit{interface} facilitated scalability and institutional adoption. The results indicate suitability for academic networks, with centralized control, synchronization with the network infrastructure, and proactive threat detection. Despite the bottleneck in API processing due to virtualization and latency, the system demonstrated consistent performance and represents an accessible alternative for institutions seeking security policies aligned with the \textit{Zero Trust} model.

\section*{Future Work}  
The next step is to advance in the scalability analysis to assess the system’s behavior as the number of devices increases, taking into account resource consumption, latency, incident flow, and potential bottlenecks. Stress tests and simulations with a high number of connections will be essential to validate the robustness of the solution in large institutional environments and to enable fine-tuning of the architecture for scenarios with hundreds or thousands of devices.

Considering the average time of 28.6 seconds between the start of the attack and its complete blocking, there are plans to investigate strategies to reduce this interval, such as optimizations in communication between services, improvements in asynchronous processing, and mechanisms for device pre-validation, in order to reduce the exposure window and mitigate the risk of unauthorized access. In addition, there is a prospect of developing mechanisms for secure information sharing between institutions, integrating advanced threat detection techniques supported by AI and machine learning, and expanding the architecture to new contexts, including smart city networks and industrial applications aligned with Industry 4.0.

\section*{Acknowledgments}
This research received partial support from the National Education and Research Network (RNP), through the IoTEdu Working Group\footnote{\url{https://gt-iotedu.github.io}}, and from the Research Support Foundation of the State of Rio Grande do Sul (FAPERGS), through grant agreements 24/2551-0001368-7 and 24/2551-0000726-1.

\bibliographystyle{ieeetr}
\bibliography{referencias}

\end{document}